\documentclass[prl,twocolumn,aps,showpacs,amsfonts,amsmath,amssymb,floatfix]
{revtex4}
\usepackage[dvips]{graphicx}

\newcommand{\be}{\begin{equation}}
\newcommand{\ee}{\end{equation}}
\begin{document}
\title{Exceeding classical capacity limit in quantum optical channel}
\author{Mikio Fujiwara}
\author{Masahiro Takeoka}
\author{Jun Mizuno}
\author{Masahide Sasaki}
\email{psasaki@crl.go.jp}
\affiliation{%
     Communications Research Laboratory,
     Koganei, Tokyo 184-8795, Japan}
%
%
\begin{abstract}

The amount of information transmissible through a communications 
channel is determined 
by the noise characteristics of the channel 
and 
by the quantities of available transmission resources. 
In classical information theory, 
the amount of transmissible information can be increased 
twice at most 
when the transmission resource 
(e.g. the code length, the bandwidth, the signal power)  
is doubled for fixed noise characteristics. 
%
In quantum information theory, however, 
the amount of information transmitted can increase 
even more than twice. 
We present a proof-of-principle demonstration of 
this super-additivity of classical capacity of a quantum channel 
by using the ternary symmetric states of a single photon, and 
by event selection from a weak coherent light source. 
We also show how the super-additive coding gain, 
even in a small code length, 
can boost the communication performance 
of conventional coding technique. 

\end{abstract}
\pacs{03.67.Hk, 03.65.Ta, 42.50.--p}


%
\date{\today}
\maketitle

In any transmission of signals at the quantum level, 
such as a long-haul optical communication 
where the signals at the receiving end are weak coherent pulses, 
ambiguity among signals may be more a matter of non-commutativity 
of quantum states, 
i.e. $\hat{\rho}_0 \hat{\rho}_1 \ne \hat{\rho}_1 \hat{\rho}_0$ 
rather than any classical noise. 
Such states can never be distinguished perfectly even in principle. 
This imposes an inevitable error in signal detection 
even in an ideal communications system. 
It was only recently that communication theory was extended into 
quantum domain to include this aspect of ambiguity, 
and 
the expressions of channel capacity 
were finally obtained 
\cite{Q_capacity_theorems}. 
Classical communication theory 
\cite{Shannon48_Gallager_book}
describes the special case of the signals 
prepared in commuting density matrices.

For reliable transmission in the presence of noise, 
redundancy must be introduced in representing messages by letters, 
such as $\{0,1\}$, 
so as to correct errors at the receiving side. 
The capacity is associated with the functional meaning of 
this channel coding. 
Messages of $k$ [bit] are encoded into 
block sequences of given letters in length $n$ $(>k)$. 
The $n-k$ [bit] redundancy allows one 
to correct errors at the receiving side. 
For a channel with a capacity $C$ [bit/letter], 
it is possible 
\cite{Shannon48_Gallager_book} 
with the rate $R=k/n\,<\,C$ to reproduce $k$ bit messages 
with an error probability as small as desired 
by appropriate encoding and decoding in the limit $n \to \infty$.

In extending the theory of capacity into quantum domain, 
primary concern is decoding of codewords made of 
non-commuting density matrices of letters. 
This is a non-trivial problem of quantum measurement. 
Actually, 
the optimal decoding essentially uses a process of entangling 
letter states constituting codewords prior to the measurement 
to enhance the distinguishability of signals. 
Such a process is nothing but 
a quantum computation on codeword states. 
This is a new aspect, 
not found in conventional coding techniques, 
and leads to a larger capacity. 
A significant consequence of 
this so called quantum collective decoding, 
is that 
the capacity can increase even more than twice 
when the code length is doubled. 
In classical information theory, on the contrary, 
the capacity can be increased twice at most. 
This feature, the super-additive quantum coding gain 
\cite{Holevo79_QuantCap,PeresWootters91,Sasaki98a,
Buck00_SupAdd,Usami01_SupAdd}, 
will be an important design rule 
for communications at the quantum level.

The theory of capacity, 
however, 
generally gives no guidance on 
how to construct codes that approach the capacity. 
The practical problem is then 
to find good codes for small blocks. 
Although several coding schemes have been proposed 
to exhibit super-additive coding gain
\cite{PeresWootters91,Sasaki98a,Buck00_SupAdd,Usami01_SupAdd}, 
little attention has been paid to this topic so far, 
and no experimental work has been reported yet. 
In fact, putting these theoretical predictions into practice 
has been considered as a formidable task with present technologies. 
In this letter, 
we experimentally demonstrate the super-additive coding gain 
by designing a coding circuit for a quantum channel 
consisting of the ternary symmetric states 
in a two-state system (qubit) of a single photon.

For binary non-orthogonal pure states, 
the most basic signals, 
the super-additive coding gain is predicted 
\cite{Buck00_SupAdd} 
for the minimum length, $n=2$. 
The amount of gain, 
however, 
is so small to be observed experimentally, 
that is, 
5.2$\times10^{-4}$ [bit] 
as the net increase of retrievable information per letter 
from the classical limit. 
For $n=3$, 
the net gain of 0.009 [bit] is predicted 
\cite{Sasaki98a}, 
however, 
this requires quantum gating more than ten steps 
with high precision, 
which is something hard to do. 
Therefore we consider the letter state set 
that shows the largest amount of the coding gain 
with the minimum code length, $n=2$, 
among the known codes 
\cite{Sasaki98a,Buck00_SupAdd,Usami01_SupAdd}.

Let us consider the set of the ternary letters 
$\{0,\,1,\,2\}$ 
conveyed by the symmetric states of a qubit system, 
$\hat{\rho}_x = \vert\psi_x\rangle\langle\psi_x\vert$ 
with 
$\vert \psi_0 \rangle = \vert 0 \rangle$,
$\vert \psi_1 \rangle = - \frac{1}{2} \vert 0 \rangle 
	               - \frac{\sqrt{3}}{2} \vert 1 \rangle$,
$\vert \psi_2 \rangle = - \frac{1}{2} \vert 0 \rangle 
		       + \frac{\sqrt{3}}{2} \vert 1 \rangle$.
%
%
%
%
Here $\{|0\rangle,\,|1\rangle\}$ is the orthonormal basis set. 
We assume that 
these states arrive at the receiver's hand 
through a noiseless transmission line. 
If the letter states were prepared in commuting density matrices, 
they could be distinguished perfectly, 
and $\log_2 3$ [bit] of information 
(the maximum Shannon entropy of the set $\{0,\,1,\,2\}$) 
could be faithfully retrieved per letter, 
meaning that 
the capacity would be $\log_2 3$ [bit/letter]. 
However, 
the states $\hat{\rho}_x$ here 
are non-commuting, 
and distinguishing them is always associated with finite errors. 
In fact, 
the average error probability can never be lower than 1/3 
when they are used with equal prior probabilities 
\cite{Helstrom_QDET}.

The capacity is matheatically given 
\cite{Shannon48_Gallager_book} 
based on 
the mutual information $I(X:Y)$ which is defined from 
the input variable $X=\{x\}$, 
the output variable $Y=\{y\}$, 
the prior distribution $\{P(x)\}$ of $X$ 
and 
the conditional probability $\{P(y|x)\}$ of $Y$ 
for given $X$. 
For the given channel model [$P(y|x)$], 
the capacity is defined by $C=\max_{\{P(x)\}}I(X:Y)$. 
In the quantum context, 
on the other hand, 
only the input variable $X$ 
and the corresponding set of quantum states $\{\hat{\rho}_x\}$ 
at the receiver's hand are given. 
The output variable $Y$ is to be sought 
for the best quantum measurement, 
i.e. a POVM $\{ \hat{\Pi}_y \}$. 
The channel matrix elements are now given by 
$P(y|x)={\rm Tr}(\hat{\Pi}_y \hat{\rho}_x)$, 
and one is to find the optimized quantity 
\cite{Mizuno02_ImaxExp}
\begin{equation}
C_1=\max_{\{P(x)\}}\max_{\{\hat\Pi_y\}} I(X:Y). 
\label{C1_def}
\end{equation}
For the ternary set $\{\vert\psi_x\rangle\}$, 
the $C_1$ was evaluated 
as $0.6454$ [bit/letter] 
which is attained by using only two of the three letters, 
say $\{|\psi_0\rangle,\,|\psi_1\rangle\}$, 
with equal probability 1/2 
and by applying the binary measurement 
to form a binary symmetric channel 
\cite{Osaki_Shor_C1}. 
%
%
The quantity $C_1$ is, 
however, 
not the ultimate capacity allowed by quantum mechanics. 
In fact $C_1$ specifies the classical capacity limit 
when the given initial channel
%
%
%
%
is used with classical channel coding 
\cite{FujiwaraNagaoka98}. 
It is this quantity 
that limits the performance of 
all conventional communications systems.

\begin{figure}
\begin{center}
\includegraphics[width=0.33\textwidth]{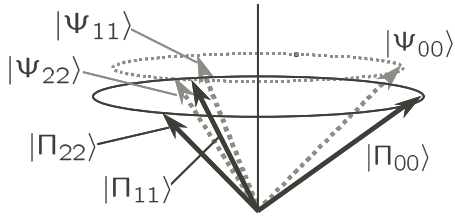}
\end{center}
\caption{\label{fig:3D_rep}
Geometrical representation of the codeword (dotted arrows) 
and decoding (solid arrows) state vectors
in a real three dimensional space. 
}
\end{figure}

Now let us consider a quantum channel coding of length two. 
There are nine possible sequences 
in length two coding of three letters. 
Peres and Wootters showed \cite{PeresWootters91} 
that $I(X^2:Y^2)=1.3690$ [bit] of information can be retrieved 
in principle, 
which is greater than 
twice of the classical limit $2C_1=1.2908$ [bit]. 
This can be achieved in the following way; 
only three sequences  
$\vert\Psi_{xx}\rangle=\vert\psi_x\rangle\otimes\vert\psi_x\rangle$ 
$(x=0,\,1,\,2)$ 
are used as the codewords with equal probability 1/3, 
and 
they are decoded by the measurement represented 
by the elements 
$\hat{\Pi}_{yy} = \vert\Pi_{yy}\rangle\langle\Pi_{yy}\vert$ 
$(y=0,\,1,\,2)$ which compose the orthonormal basis 
expanding $\{\vert\Psi_{xx}\rangle\}$, that is, 
\begin{subequations}
\label{ternary_state_POVM}
\begin{eqnarray}
| \Psi_{00} \rangle & = & 
      c \,\, | \Pi_{00} \rangle
    - \frac{s}{\sqrt{2}} \,\, | \Pi_{11} \rangle
    - \frac{s}{\sqrt{2}} \,\, | \Pi_{22} \rangle, 
\\
| \Psi_{11} \rangle & = & 
    - \frac{s}{\sqrt{2}} \,\, | \Pi_{00} \rangle
    + c \,\, | \Pi_{11} \rangle
    - \frac{s}{\sqrt{2}} \,\, | \Pi_{22} \rangle,
\\
| \Psi_{22} \rangle & = & 
    - \frac{s}{\sqrt{2}} \,\, | \Pi_{00} \rangle
    - \frac{s}{\sqrt{2}} \,\, | \Pi_{11} \rangle
    + c \,\, | \Pi_{22} \rangle, 
\end{eqnarray}
\end{subequations}
where 
$c={\rm cos}{\gamma\over2}=(\sqrt{2}+1)/\sqrt{6}$, 
and 
$s={\rm sin}{\gamma\over2}=(\sqrt{2}-1)/\sqrt{6}$ 
($\gamma\simeq19.47^\circ$). 
Fig.~\ref{fig:3D_rep} shows a geometrical representation of 
Eq.~(\ref{ternary_state_POVM}). 
%
%
%
%
The super-additive coding gain is 
$I(X^2:Y^2)/2-C_1=0.0391$ [bit/letter].

%
%
%
%
To demonstrate this gain, 
we must be able to entangle two letter states 
at the receiver's hand prior to a measurement. 
Unfortunately quantum gating operations demonstrated to date 
are not precise enough 
to observe the small quantum coding gain.
%
%
%
%
Therefore 
our method for proof-of-principle demonstration is based on 
the use of two physically different kinds of qubits 
of a single photon. 
The first and second letters of a codeword are drawn from 
the ternary letter state sets made of a polarization 
and a location qubits, respectively. 
Then entangling the polarization and location degrees of freedom 
of a photon can be performed by linear optical components 
with very high accuracy. 
The polarization qubit consists of 
the horizontal $|H\rangle$ 
and vertical $|V\rangle$ 
polarization states of a single photon. 
%
%
The location qubit for the second letter is realized 
by guiding the polarization qubit into 
two different paths A and B 
through a polarizing beamsplitter (PBS) 
which reflects the vertical polarization 
and transmits the horizontal polarization.
%
%
%
%
Thus, 
the length two coding space is spanned 
by the two orthonormal basis sets 
\cite{Cerf_Spreeuw_Takeuchi_LinearOpt}
\begin{subequations}
\label{eq-3-3}
\begin{eqnarray}
|00\rangle & = & |0\rangle_P \otimes |0\rangle_L 
	        = |H\rangle_A \otimes |{\rm vacuum}\rangle_B,
\\
|01\rangle & = & |0\rangle_P \otimes |1\rangle_L 
		= |{\rm vacuum}\rangle_A \otimes |H\rangle_B,
\\
|10\rangle & = & |1\rangle_P \otimes |0\rangle_L 
		= |V\rangle_A \otimes |{\rm vacuum}\rangle_B,
\\
|11\rangle & = & |1\rangle_P \otimes |1\rangle_L 
		= |{\rm vacuum}\rangle_A \otimes |V\rangle_B.
\end{eqnarray}
\end{subequations}
The codeword states 
$|\Psi_{xx}\rangle
=|\psi_x\rangle_P \otimes |\psi_x\rangle_L$ 
are represented in this product space.   
%
%
%
%
Thus 
the increase in resources in our coding format is due to 
doubling the spatial resource 
which is analogous to doubling the transmission bandwidth, 
as opposed to doubling the number of polarized photons. 
We want then to observe the increase 
more than double the amount of information transmitted. 

\begin{figure}
\begin{center}
\includegraphics[width=0.46\textwidth]{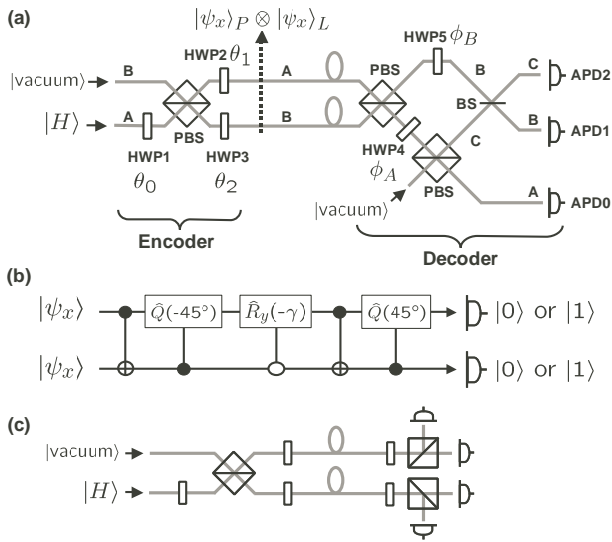}
\end{center}
\caption{\label{fig:circuit}
(a) Encoding and decoding circuits. 
The angles of the HWPs, 
$\theta_0$, $\theta_1$, and $\theta_2$ 
are chosen as described in the text, 
and 
$\phi_A=-\gamma/2=-9.74^{\circ}$ 
and 
$\phi_B=-45^{\circ}$. 
(b) Quantum circuit to realize the collective decoding 
by $\{|\Pi_{yy}\rangle\}$, 
which can be applied to any physical qubits. 
A received codeword state is first transformed 
by the five controlled gates, 
and 
is then detected by a standard von Neumann measurement 
on each letter. 
The open circle indicates conditioning 
on the control qubit being set to zero, 
and 
$\hat{Q}(\varphi) = \hat{R}_y(\varphi) \hat{\sigma}_z$, 
and $\gamma=19.47^{\circ}$. 
Other nomenclature follows 
the Ref.~\cite{Barenco95_Gates}. 
(c) Circuit for separable (classical) decoding for $C_1$. 
}
\end{figure}

An optical circuit for 
this coding 
is shown in 
Fig.~\ref{fig:circuit}(a).
The angles of the three half waveplates 
(HWPs) 
$\theta$'s are chosen as 
$(\theta_0,\,\theta_1,\,\theta_2)$
=(0$^{\circ}$,\,0$^{\circ}$,\,0$^{\circ}$), 
(30$^{\circ}$,\,$-30^{\circ}$,\,$-15^{\circ}$), and 
(30$^{\circ}$,\, 30$^{\circ}$,\, 15$^{\circ}$) for 
$|\Psi_{00}\rangle$, $|\Psi_{11}\rangle$, and 
$|\Psi_{22}\rangle$, respectively.
This decoding circuit is derived with slight modifications from 
a general circuit design 
of Fig.~\ref{fig:circuit}(b), 
which can be applied to any physical qubits. 
%
%
%
%
The received codeword is decided to be either of 
$|\Psi_{00}\rangle$, 
$|\Psi_{11}\rangle$, or 
$|\Psi_{22}\rangle$ 
according to the detection of a photon 
by the avalanche photodetector 
APD0, 
APD1, or 
APD2, 
respectively.

In our experiment, 
the CW light from a He-Ne laser at the wavelength of 632.8\,nm 
with 1\,mW power is strongly attenuated such that 
about $10^{-2}$ photons exist on average in the whole circuit. 
%
%
%
%
The signal photons are guided to the Si APDs 
whose quantum efficiency and dark count are typically 
70\% and 100 [count/sec], respectively, 
through a multimode optical fiber 
with coupling efficiency of about 80\%. 
In this setup, 
the mutual information is evaluated 
by constructing the 3-by-3 channel matrix 
$[P(yy|xx)\equiv|\langle\Pi_{yy}|\Psi_{xx}\rangle|^2]$ 
from a statistical data of single-photon events 
detected by either of the three APDs 
conditioned on an input codeword $|\Psi_{xx}\rangle$. 
The mutual information thus obtained measures the ratio of 
number of bits retrieved per number of total photon counts. 
This allows us to simulate communications of ``pure'' codeword states 
of two letters by sending and detecting the photons one by one 
through the channel. 
The error performance is then determined only by 
the non-commutativity of the signal states, 
imperfect alignment of the whole interferometer, 
and 
the dark count of the APDs.

Each polarization Mach-Zehnder interferometer 
must be adjusted simultaneously 
at a proper operating point.
This is done by using a bright reference beam and Piezo transducers 
with low noise voltage sources.
The visibility of the whole interferometer is typically 98\%. 
Once the circuit is adjusted, 
the reference beam is shut off. 
The signal light is then guided into the encoder and decoder. 
Photon counts are measured for five-second duration. 
This procedure is repeated for each codeword, 
composing a full sequence of measuring the channel matrix. 
The temporal stability in this sampling mode corresponds to 
the relative path length change within 3\,nm 
for at least more than 200\,sec, 
which causes the error in mutual information within $\pm$0.005 [bit].

\begin{figure}
\begin{center}
\includegraphics[width=0.40\textwidth]{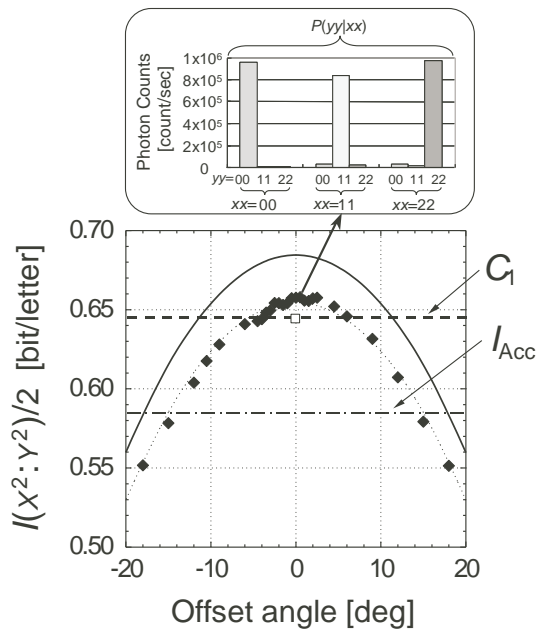}
\end{center}
\caption{\label{fig:MeasuredData} 
The upper: 
Histogram of photon counts for the channel matrix elements 
$P(yy|xx)$ 
corresponding to the maximum mutual information. 
The lower: 
Measured (diamonds) and theoretical (solid curve) 
mutual information 
as a function of the offset angle of the codeword state set 
$\{|\Psi_{xx}\rangle\}$ 
from the decoder state set 
$\{|\Pi_{yy}\rangle\}$ 
around the vertical axis in Fig.~\ref{fig:3D_rep}. 
The dotted curve is just the guide for eyes. 
The theoretical $C_1$ and 
accessible information $I_\mathrm{Acc}$ 
are shown by the dashed and one-dotted lines, 
respectively
\cite{Osaki_Shor_C1}. 
The square represents the experimental $C_1$. 
}
\end{figure}

An example of the channel matrix measured is shown as 
a histogram in 
Fig.~\ref{fig:MeasuredData}. 
Ideally, 
the diagonal and off-diagonal elements must be 
$c^2=$0.9714 and $s^2/2=$0.0143, respectively. 
The total events counted for 1 sec is of order $10^6$, 
while the average count for the off-diagonal elements is 
about $1.9\times10^4$. 
The background photons amount to about 300 [count/sec]. 
Including dark counts, 
the total background photon count is 2\% of the average count 
for the off-diagonal elements. 
The mutual information is evaluated as 
$I(X^2:Y^2)=1.312\pm0.005$ [bit]. 
For experimental clarity, 
we measured the variation of the mutual information 
when the codeword state set $\{|\Psi_{xx}\rangle\}$ is rotated 
with respect to the decoder state set $\{|\Pi_{yy}\rangle\}$ 
around the vertical axis in 
Fig.~\ref{fig:3D_rep}. 
The result is shown in 
Fig.~\ref{fig:MeasuredData}. 
The gap between the data points (diamonds) 
and the ideal curve (solid curve) 
is mainly attributed to the imperfection of the PBSs. 
Fluctuation of the data points are mainly due to thermal drifts. 
The corresponding error bars ($\sim\pm$0.005 [bit]) are about 
the same size of the diamonds.  
The measured mutual information per letter, 
$0.656\pm0.003$ [bit/letter], 
is clearly greater than the classical theoretical limit 
$C_1=0.6454$ [bit/letter], 
which is the level shown by the dashed lines. 
The white square represents the experimental 
$C_1$, 
$0.644\pm0.001$ [bit/letter]. 
This is measured 
by the circuit for classical decoding of 
Fig.~\ref{fig:circuit}(c), 
which does not entangle the polarization and location qubits. 
%
%
The retrieved information can never exceed $2C_1$. 
Our results clearly show that 
when an appropriate quantum circuit for entangling the letter states 
is inserted in front of the separable decoding, 
one can retrieve information more than twice per letter. 


The super-additive coding gain in small blocks 
is not only valuable as a proof-of-principle demonstration 
but also of practical importance in quantum-limited communications. 
Even a two-qubit quantum circuit like 
Fig.~\ref{fig:circuit}(b) 
is useful in boosting the performance of a classical decoder. 
It can be shown that the decoding error can be greatly reduced 
by inserting the quantum circuit in front of the classical decoder. 
The quantum circuit processes a received codeword state 
quantum collectively 
prior to converting it into a classical signal. 
We call such a scheme quantum-classical hybrid coding (QCHC). 
The theoretical error exponents 
\cite{Shannon48_Gallager_book} 
of QCHC and all-classical coding (ACC) 
in the ternary letter-state case are listed in Table~\ref{tab:table1} 
for low and high transmission rates $R$. 
The improvement is more drastic in the higher rate limit. 
For the rate $R=0.62$ [bit/letter] (96\% of $C_1$), 
it is possible to reduce the decoding error as 
$P_e=2^{-\frac{n}{2}E(R)}=2^{-0.0488n}$ 
by an appropriate classical coding 
with the composite letters \{00, 11, 22\} 
assisted by the pair-wise quantum decoding. 
To achieve a standard error-free criterion $P_e=10^{-9}$, 
QCHC requires the code length 
$n=614$ (307 composite letter pairs), 
whereas ACC typically needs 
$n=57300$. 
As codes get longer, 
the complexity of the decoder, 
such as the total number of arithmetic operations, 
increases and eventually limits the effective transmission speed. 
For some asymptotically good codes, 
the total number of arithmetic operations is evaluated 
\cite{Hirasawa80} 
to be typically of order ($n\log\,n)^2$. 
Then the reduction of code length attained by QCHC 
will be practically significant 
in the trade-off between performance and decoding complexity.
This suggests a useful application of small scale quantum computation. 

\begin{table}
\caption{\label{tab:table1}
Error exponent $E(R)$ of QCHC and ACC.}
\begin{ruledtabular}
\begin{tabular}{lll}
$R$ [bit/letter] & $E(R)$ of QCHC & $E(R)$ of ACC \\
\hline
0.1 & 0.842 & 0.315 \\
0.62 & 9.753$\times10^{-2}$ & 5.218$\times10^{-4}$ \\
\end{tabular}
\end{ruledtabular}
\end{table}

%
%
%
%

The super-additive quantum coding gain 
should eventually be applied to more practical resources 
such as optical pulses of coherent states, 
especially, heavily attenuated coherent states 
$\{|\alpha_k \rangle\}$ 
of phase-shift and/or amplitude-shift keying. 
Unlike our channel model of single photon polarization 
and location modes, 
one must be able to entangle weak coherent pulses 
with respect to the degrees of phase and/or amplitude. 
This is another big challenge. 

%
%

%

The authors acknowledge J.~A.~Vaccaro for his valuable comments.
This work was supported by the CREST of JST.

\end{document}